\documentclass{iau}
\usepackage{graphicx}

\title[Diffuse gamma-ray emission] 
{Diffuse gamma-ray emission from the Galactic center and implications of its past activities}

\author[Yutaka Fujita, Shigeo S. Kimura, \& Kohta Murase]   
{Yutaka Fujita$^1$, Shigeo S. Kimura$^2$
 \and Kohta Murase$^3$}

\affiliation{$^1$Department of Earth and Space Science, Graduate School of Science, \\ Osaka University,
Toyonaka, Osaka 560-0043, Japan \\ email: {\tt fujita@vega.ess.sci.osaka-u.ac.jp} \\[\affilskip]
$^2$Frontier Research Institute for Interdisciplinary Sciences, Tohoku University, \\ Sendai 980-8578, Japan
and Astronomical Institute, \\ Tohoku University, Sendai 980-8578, Japan
\\[\affilskip]
$^3$Department of Physics; Department of Astronomy \& Astrophysics; \\ Center for Particle and Gravitational Astrophysics, \\ The Pennsylvania State University, University Park, PA 16802, USA}

\pubyear{2016}
\volume{322}  
\jname{The Multi-Messenger Astrophysics of the Galactic Centre}
\editors{A.C. Editor, B.D. Editor \& C.E. Editor, eds.}
\begin{document}

\maketitle

\begin{abstract}
It has been indicated that low-luminosity active galactic nuclei
(LLAGNs) are accelerating high-energy cosmic-ray (CR) protons in their
radiatively inefficient accretion flows (RIAFs). If this is the case,
Sagittarius~A* (Sgr~A*) should also be generating CR protons, because
Sgr~A* is a LLAGN. Based on this scenario, we calculate a production
rate of CR protons in Sgr~A* and their diffusion in the central
molecular zone (CMZ) around Sgr~A*. The CR protons diffusing in the CMZ
create gamma-rays through $pp$ interaction. We show that the gamma-ray
luminosity and spectrum are consistent with observations if Sgr~A* was
active in the past. 
\end{abstract}

\firstsection 
\section{Introduction}

The IceCube Collaboration reported the detection of extraterrestrial
neutrinos (\cite[Aartsen et al. 2013]{Aartsen_etal13}; \cite[Aartsen et
al. 2014]{Aartsen_etal14}; \cite[Aartsen et
al. 2015]{Aartsen_etal15}). Although their origin is not understood, the
uniform arrival directions of the neutrinos indicate their extragalactic
origin. Recently, \cite[Kimura et al. (2015)]{Kimura etal15} proposed
that the neutrinos are coming from low-luminosity active galactic nuclei
(LLAGNs). Those AGNs are thought to have radiatively inefficient
accretion flows (RIAFs), in which gas is hot and tenuous. In this
environment, thermalization of gas particles is inefficient and some of
them can be stochastically accelerated. The cosmic ray (CR) protons
accelerated in this process interact with photons ($p\gamma$
interaction) and other protons ($pp$ interaction), and produce
gamma-rays and neutrinos.

Sagittarius~A* (Sgr~A*) is the AGN of the Galaxy, and it is known as a
LLAGN. This means that CR protons may be accelerated in the RIAF of
Sgr~A* and they may be injected in the interstellar space. Sgr~A* is
surrounded by a huge amount of dense molecular gas called the central
molecular zone (CMZ). The mass and size of the CMZ is $M_{\rm CMZ}\sim
10^7\: M_\odot$ and $R_{\rm CMZ}\sim 100$~pc, respectively
(e.g. \cite[Morris \& Serabyn 1996]{Morris96}). If Sgr~A* is actually
producing CR protons, some of them should enter the CMZ, and generate
gamma-rays and neutrinos through $pp$ interaction with protons in the
CMZ. In fact, gamma-rays have been observed from the CMZ with the High
Energy Stereoscopic System (HESS; \cite[Aharonian et
al. 2006]{Aharonian_etal06}; \cite[H.E.S.S. collaboration
2016]{HESS16}). In this study, we consider the CR acceleration in the
RIAF of Sgr~A* and the diffusion of the accelerated protons in the
CMZ. We investigate the gamma-ray and neutrino production in the CMZ. We
compare the predicted gamma-ray emission with observations. The details
of this study has been published in \cite[Fujita et
al. (2015)]{Fujita15}.

\section{CR proton acceleration in Sgr~A*}

Since we expect that the CRs are accelerated in a small region around
the supermassive black hole (SMBH) in Sgr~A*, we adopt an one-zone model
for the acceleration (\cite[Kimura et al. 2015]{Kimura_etal15}). The
typical energy of the CR protons can be estimated by equating their
acceleration time to their escape time from the RIAF. The result is
\begin{eqnarray}
\label{eq:geq}
 \frac{E_{p,\rm eq}}{m_p c^2} &\sim& 1.4\times 10^5 
\left(\frac{\dot{m}}{0.01}\right)^{1/2}
\left(\frac{M_{\rm BH}}{1\times 10^7\: M_\odot}\right)^{1/2}\nonumber\\
& &\times \left(\frac{\alpha}{0.1}\right)^{1/2}
\left(\frac{\zeta}{0.1}\right)^3
\left(\frac{\beta}{3}\right)^{-2}
\left(\frac{R_{\rm acc}}{10\: R_S}\right)^{-7/4}\:,
\end{eqnarray}
where $m_p$ is the proton mass, $\dot{m}$ is the normalized accretion
rate $\dot{m}=\dot{M}/\dot{M}_{\rm Edd}$, $\dot{M}_{\rm Edd}$ is the
Eddington accretion rate, $\alpha$ is the alpha parameter of the
accretion flow, $\zeta$ is the ratio of the strength of turbulent fields
to that of the non-turbulent fields, $\beta$ is the plasma beta
parameter, $R_{\rm acc}$ is the typical radius where particles are
accelerated, and $R_S$ is the Schwarzschild radius of the SMBH. As
fiducial parameters, we adopt $\alpha=0.1$, $\zeta=0.05$, $\beta=3$, and
$R_{\rm acc}=10\: R_S$ to be consistent with the IceCube observations
(\cite[Kimura et al. 2015]{Kimura_etal15}). The proton luminosity of the
RIAF is given by $L_{p,\rm tot}=\eta_{\rm cr} \dot{M}c^2$, where
$\eta_{\rm cr}$ is the parameter and we take $\eta_{\rm cr}=0.015$ as
the fiducial value. Assuming that the particles are accelerated
stochastically, the functional form of the CR spectrum is
\begin{equation}
\label{eq:sp}
 \dot{N}(x)dx \propto x^{(7-3q)/2}K_{(b-1)/2}(x^{2-q})dx\:,
\end{equation}
where $x=p/p_{\rm cut}$, $K_\nu$ is the Bessel function, and $b=3/(2-q)$
(\cite[Becker et al. 2006]{Becker_etal06}). The cut-off momentum is
given by $p_{\rm cut}=(2-q)^{1/(2-q)}p_{\rm eq}=p_{\rm eq}/27$, where
$p_{\rm eq}=E_{p,{\rm eq}}/c$ and $q=5/3$ (Kolmogorov type
turbulence). Normalization of Eq.~(\ref{eq:sp}) can be obtained by
setting that the total proton luminosity is $L_{p,\rm tot}$.
 
\section{Diffusion of Protons in the CMZ}

Since we are only interested in the diffusion of CRs in the direction of
the CMZ or the Galactic plane, we solve a spherically symmetric
diffusion equation for the sake of simplicity:
\begin{equation}
\label{eq:diff}
 \frac{\partial f}{\partial t}
= \frac{1}{r^2}\frac{\partial}{\partial r}\left(r^2 \kappa\frac{\partial
					       f}{\partial r}\right)
+ Q\:.
\end{equation}
We only consider the diffusion inside the CMZ ($r<R_{\rm CMZ}$) and
steady state solutions ($\partial f/\partial t=0$). The source term is
written as $\int 4\pi c p^3 Q dp =\lambda L_{p,\rm tot} =
\lambda\eta_{\rm cr} \dot{M}c^2$, where $\lambda$ is the effective
covering factor, which is expected to be $\lambda\ll 1$. The diffusion
coefficient is
\begin{equation}
\label{eq:diffc}
 \kappa = 10^{28} \left(\frac{E_p}{10\rm\; GeV}\right)^{0.5}
\left(\frac{B}{3\rm\: \mu G}\right)^{-0.5} \rm\: cm^2 s^{-1}\:,
\end{equation}
where $E_p$ is the particle energy (\cite[Gabici et
al. 2009]{Gabici_etal09}). We assume that the magnetic field in the CMZ
is $B=1$~mG (\cite[Morris \& Serabyn 1996]{Morris96}). The CR protons
interact with protons in the CMZ and generate gamma-ray photons and
neutrinos. We calculate their production rates using the models of
\cite{Karlsson08} and \cite{Kelner_etal06}.

\section{Results}

\begin{figure}[t]
\begin{center}
 \includegraphics[width=4.0in]{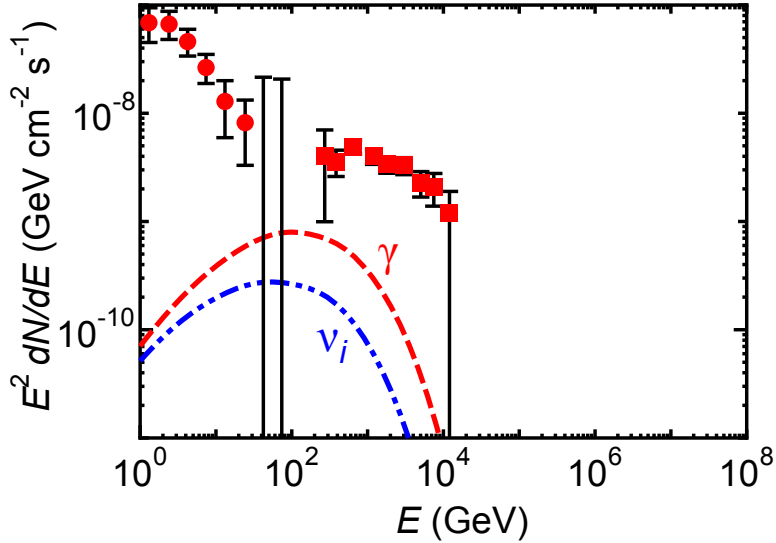} 
 \caption{Predicated gamma-ray flux (dashed line) and neutrino flux
(two-dot dashed line) from the CMZ when $\dot{m}=4.2\times 10^{-6}$ and
$\lambda=0.01$. Filled circles and squares are the Fermi and HESS
observations, respectively (\cite[Yusef-Zadeh et
al. 2013]{Yusef-Zadeh_etal13}; \cite[Aharonian et
al. 2006]{Aharonian_etal06}). This figure is cited from \cite[Fujita et
al. (2015)]{Fujita15}.}  \label{fig1}
\end{center}
\end{figure}

We assume that the typical size of the CMZ is $R_{\rm CMZ}=130$~pc, and
the density is $\rho_{\rm CMZ}=1.4\times 10^{-22}\rm\: g\:cm^{-3}$. The
mass of the SMBH is $M_{\rm BH}=4.3\times 10^6\: M_\odot$
(\cite[Gillessen et al. 2009]{Gillessen_etal09}).  Fig.~\ref{fig1} shows
the gamma-ray and neutrino fluxes from the CMZ when the accretion rate
is the currently observed one or $\dot{m}=4.2\times 10^{-6}$ (\cite[Yuan
et al. 2003]{yuan_etal03}), and when the effective covering factor is
$\lambda=0.01$. Apparently, the predicated gamma-ray flux underestimates
the Fermi and HESS observations.

\begin{figure}[t]
\begin{center}
 \includegraphics[width=4.0in]{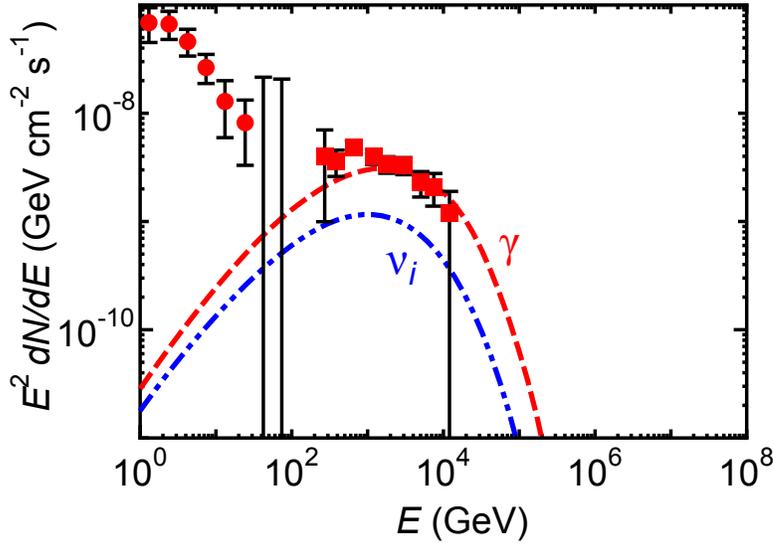} 
 \caption{Same as Fig.~\ref{fig1} but for $\dot{m}=0.001$, and
 $\lambda=5\times 10^{-4}$. This figure is cited from \cite[Fujita et
 al. (2015)]{Fujita15}.}  \label{fig2}
\end{center}
\end{figure}

However, observations of X-ray echos have shown that Sgr~A* was much
more active $\gtrsim 100$~yrs ago and the accretion rate was
$10^3$--$10^4$ times as much as the current one (\cite[Koyama et
al. 1996]{Koyama_etal96}; \cite[Ryu et al. 2013]{Ryu_etal13}). Thus, we
calculate the gamma-ray and neutrino fluxes when $\dot{m}=10^{-3}$ and
$\lambda=5\times 10^{-4}$, and show the results in Fig.~\ref{fig2}. The
decline of $\dot{m}$ for the recent $\sim 100$~yrs does not affect the
results. Fig~\ref{fig2} indicates that our prediction is consistent with
TeV gamma-ray observations with HESS. However, our model cannot explain
the GeV gamma-ray observations with Fermi. This may mean that the origin
of the GeV emission is different from that of the TeV emission. The
neutrinos could be observed with KM3Net in the future.

In summary, we have shown that the TeV gamma-ray emission from the CMZ
can be explained if Sgr~A* accelerates CR protons in the RIAF and if it
was more active in the past. In fact, recent HESS observations have
suggested that the source of CRs responsible for the gamma-rays from the
CMZ is located at the Galactic center (\cite[H.E.S.S. collaboration
2016]{HESS16}). It is worth noting that Sgr~A* can accelerate PeV CRs in
our model. Thus, Sgr~A* can also be a candidate of PeVatron in the
Galaxy.

\begin{thebibliography}{}

\bibitem[Aartsen \etal\ (2013)]{Aartsen_etal13}
{Aartsen, M.G. et al.} 2013,
\textit{Phys. Rev. Lett.}, 111, 021103

\bibitem[Aartsen \etal\ (2013)]{Aartsen_etal14}
{Aartsen, M.G. et al.} 2014,
\textit{Phys. Rev. Lett.}, 113, 101101

\bibitem[Aartsen \etal\ (2013)]{Aartsen_etal15}
{Aartsen, M.G. et al.} 2015,
\textit{Phys. Rev. D}, 91, 022001

\bibitem[Kimura, K. Murase \& K. Toma (2015)]{Kimura_etal15}
{Kimura, S.S., Murase, K., \& Toma, K.} 2015, 
\textit{ApJ}, 806, 159

\bibitem[Morris \& Serabyn (1996)]{Morris96}
{Morris. M. \& Serabyn, E.} 1996, 
\textit{ARAA}, 34, 645

\bibitem[Aharonian et al. (2006)]{Aharonian_etal06}
{Aharonian, F. et al.} 2006, 
\textit{Nature}, 439, 695

\bibitem[H.E.S.S. collaboration (2016)]{HESS16}
{H.E.S.S. collaboration} 2016, 
\textit{Nature}, 531, 476

\bibitem[Fujita, Kimura \& Murase (2015)]{Fujita15}
{Fujita, Y., Kimura, S.S., \& Murase, K.} 2015,
\textit{Phys. Rev. D}, 92, 023001 

\bibitem[Becker, Le \& Dermer (2006)]{Becker_etal06}
{Becker, P.A., Le, T., \& Dermer, C.D.} 2000, 
\textit{ApJ}, 647, 539

\bibitem[Gabici, Aharonian \& Casanova (2009)]{Gabici_etal09}
{Gabici, S., Aharonian, F.A., \& Casanova, S.} 2009, 
\textit{MNRAS}, 396, 1629

\bibitem[Karlsson \& Kamae (2008)]{Karlsson08}
{Karlsson, N., \& Kamae, T.} 2008, 
\textit{ApJ}, 674, 278

\bibitem[Kelner, Clayton \& Bugayov (2006)]{Kelner_etal06}
{Kelner, S.R., Aharonian, F.A., \& Bugayov, V.V.} 2006, 
\textit{Phys. Rev. D}, 74, 034018

\bibitem[Gillessen et al. (2009)]{Gillessen_etal09}
{Gillessen, S. Eisenhauer, F., Trippe, S., Alexander, T.,
Genzel, R., Martins, F, \& Ott, T.} 2008,
\textit{ApJ}, 692, 1075

\bibitem[Yuan, Quataert \& Narayan (2003)]{Yuan_etal03}
{Yuan, F., Quataert, E., \& Narayan, R.} 2003,
\textit{ApJ}, 598, 301

\bibitem[Yusef-Zadeh et al. (2013)]{Yusef-Zadeh_etal13}
{Yusef-Zadeh, F. et al.} 2013, 
\textit{ApJ}, 762, 33

\bibitem[Koyama et al. (1996)]{koyama_etal96}
{Koyama, K., Maeda, Y., Sonobe, T., Takeshima, T., Tanaka, Y.,
\& Yamauchi, S.} 1996,
\textit{PASJ}, 48, 249

\bibitem[Ryu et al. (2013)]{Ryu_etal13}
{Ryu, S.G., Nobukawa, M., Nakashima, S., Tsuru, T.G.,
Koyama, K, \& Uchiyama, H.} 2013,
\textit{PASJ}, 65, 33

\end{thebibliography}
\end{document}